\documentclass[aps, nofootinbib]{revtex4}
\usepackage[mathletters]{ucs}
\usepackage{psfrag}
\usepackage{array}
\usepackage{amstext,amsmath,amssymb,amsfonts,bbm}
\usepackage[latin1]{inputenc}
\usepackage[dvips]{graphicx}
\usepackage{epsfig}
\usepackage{subfigure}
\newcommand{\N}{\mathbb{N}}

\newcommand{\R}{\mathbb{R}}

\newcommand{\f}{\frac}

\newcommand{\SU}{\mathrm{SU}}

\def\arr{\rightarrow}

\def\sj{$\{6j\}$}

\newcommand{\be}{\begin{equation}}
\newcommand{\ee}{\end{equation}}
\newcommand{\bes}{\begin{eqnarray}}
\newcommand{\ees}{\end{eqnarray}}

\newcommand{\equa} [1] {\begin{equation} #1\end{equation}}

\newcommand{\tabl} [2] {\begin{array} {#1} #2 \end{array}}
\begin{document}

\title{\large\bf Pushing Further the Asymptotics of the 6j-symbol}

\author{Ma\" \i t\'e Dupuis}\email{maite.dupuis@ens-lyon.fr}
%\affiliation{Laboratoire de Physique, ENS Lyon, CNRS-UMR 5672, 46 All\'ee d'Italie, Lyon 69007, France}
%
\author{Etera R. Livine}\email{etera.livine@ens-lyon.fr}
\affiliation{Laboratoire de Physique, ENS Lyon, CNRS-UMR 5672, 46 All\'ee d'Italie, Lyon 69007, France}

\date{\small \today}

\begin{abstract}

In the context of spinfoam models for quantum gravity, we investigate the asymptotical behavior of the \sj-symbol at next-to-leading order. We compute it analytically and check our results against numerical calculations. The \sj-symbol is the building block of the Ponzano-Regge amplitudes for 3d quantum gravity, and the present analysis is directly relevant to deriving the quantum corrections to gravitational correlations in the spinfoam formalism.

\end{abstract}

\maketitle
%%%%%%%%%%%%%%%%%%%%%%%%%%%%%%%%%%%%%%%%%

%%%%%%%%%%%%%%%
\section{Introduction: Spinfoams and the \sj-symbol}
\label{intro}
%%%%%%%%%%%%%%%

The spinfoam formalism is an attempt to define rigourously a path integral for Quantum Gravity. Spinfoams can be interpreted from several perspectives, as a covariant history formulation for Loop Quantum Gravity describing the evolution of the spin network states, as an improved and quantized version of the Regge calculus for general relativity, as a quantization of ``almost topological" field theories, as a higher-dimensional generalization of the matrix models generating 2d surfaces. And they have been shown to be related to many other approaches to quantum gravity. The spinfoam model for 3d quantum gravity is the Ponzano-Regge model \cite{PR}, which was the first spinfoam model ever written. It has been shown to provide a consistent quantization of general relativity in three space-time dimensions (for both Riemannian and Lorentzian signatures). The main spinfoam models for 4d quantum gravity are the Barrett-Crane model \cite{BC} and the more recent EPR-FK-LS family of spinfoam models \cite{EPR,FK,LS}. They are related to the reformulation of general relativity as a constrained topological BF-theory and were mostly constructed as a discretization of the path integral over space-time geometries. We also point other spinfoam constructions based on another way to discretize the constraints reducing BF-theory to gravity \cite{reisenberger}, on the quantization of the MacDowell-Mansouri action for gravity \cite{artem}, on the group field theory approach \cite{oriti} and on a generalization of the Barrett-Crane model better suited for renormalization  \cite{morecoupling}.

Such spinfoam models provide a description of the quantum geometry of space-time at the Planck scale. The main issue is then to extract semi-classical information from the formalism and to show its relation to the more standard perturbative approach to the quantization of general relativity (as a quantum field theory). Solving this question amounts to proving that we recover general relativity in a large scale (or low energy) regime of the spinfoam models and to showing  how to compute the quantum corrections to the classical dynamics of the gravitational field.
A proposal to address this problem is the ``spinfoam graviton" framework proposed by Rovelli and collaborators \cite{graviton}. It defines the propagators and correlation functions for geometric observables, mainly the area, from which we can extract information about the (effective) space-time metric and its (quantum) fluctuations. Most explicit calculations in this framework have been done at the leading order (in the scale parameter) and for the simplest space-time triangulation (a single tetrahedron in 3d and a single 4-simplex in 4d). In order to make the link with the standard QFT perturbative expansion, we need to be able to push these calculations further and calculate the correlations both at higher order (``loop corrections") and for more refined triangulations (smoother boundary state). In the present work, we focus on the first aspect: the leading order of the correlations gives the classical propagator of the graviton and we would like to compute the higher order (quantum) corrections.
Following the lines of \cite{josh,valentin,4d}, this requires understanding the corrections to the asymptotical behavior of the spinfoam vertex amplitude, which is the amplitude associated to a single tetrahedron in 3d quantum gravity or to a single 4-simplex in 4d models. This is the basic building blocks of spinfoam models, which are then constructed by gluing these spinfoam vertices in some particular way in order to describe the whole space-time. In the Ponzano-Regge model, the spinfoam vertex is given by the \sj-symbol from the recoupling theory of the representations of $\SU(2)$. The Barrett-Crane model is defined by the $\{10j\}$-symbol and the more recent models use the EPR or FK vertex amplitudes. The present paper focuses on the \sj-symbol, relevant for 3d quantum gravity.

They are three basic ways to compute the leading order asymptotics of the \sj-symbol and show its relation to the Regge action for 3d gravity:
\begin{itemize}

\item {\it Recursion relations}~\cite{SG}: Using the invariance of the \sj-symbol under Pachner moves (Biedenharn-Elliott identity) or directly its definition as a recoupling coefficient, one can derive a recursion relation for \sj-symbol. This recursion formula is actually very useful for numerical computations, but it can also be approximated at large spins by a (second order) differential equation. One then derive the asymptotics from a WKB approximation.

\item {\it Integral formula}~\cite{freidel,barrett}: One can write the square of the \sj-symbol as an integral over four copies of $\SU(2)$. In the large spin regime, we can use saddle point techniques and one derives the right asymptotics after a careful analysis of non-degenerate and degenerate configurations for the saddle points. This is the technique used to derive the asymptotics of the Barrett-Crane and EPR-FK vertex amplitudes.

\item {\it Brute-force approximation}~\cite{razvan}: One can start from the explicit algebraic formula of the \sj-symbol as a sum over some products of factorials. Using the Stirling formula and after lengthy calculations, we approximate the sum by an integral and use saddle point techniques again which lead to the same asymptotics.

\end{itemize}
We also point out the more sophisticated and rigorous proof of the asymptotics by Roberts \cite{roberts} based on geometric quantization, but that also uses an integral formula and saddle point methods.

The present goal is to push these approaches one step further and derive the first correction to the leading order formula. As a first attempt, we focus on the third method and show how to extract the next-to-leading order corrections through a brutal approximation of the explicit algebraic expression of the \sj symbol. We compare our results with the previous calculations for the cases of the equilateral and isosceles tetrahedra \cite{valentin} and check the general case against numerical calculations. Although the final explicit formula for this next-to-leading order in the general case is not particularly pretty, we prove that it is indeed possible to compute it analytically exactly and we show that we could extract all orders of the \sj-symbol using the same procedure. This is a necessary step towards providing explicit formulas or procedures to compute all orders of the perturbative expansion (in term of the length scale) of the graviton correlations in spinfoam models.

We can also use the more subtle approach of the recursion relation. This requires a careful analysis of the recursion relation and computing the corrections to the WKB approximation \cite{next1}.
Or we could use the integral formula technique, then one should be particularly careful when dealing with the degenerate contributions to the \sj-symbol.
%\cite{next2}.

%%%%%%%%%%%%%%%
\section{The \sj-symbol}
%%%%%%%%%%%%%%%

The \sj-symbol is the basic building block of the Ponzano-Regge model which is a state sum model for 3d Euclidean gravity formulated as a $\SU(2)$ gauge theory. The Ponzano-Regge model is defined over a triangulation of space-time: we build the 3d space-time manifold from tetrahedra glued together along their respective triangles and edges. We assign an irreducible representation (irreps) of $\SU(2)$ to each edge $e$ of the triangulation. These irreps are labeled by a half-integer $j_e\in\N/2$, the spin, and the dimension of the corresponding representation space is given by $d_{j_{e}}=2j_{e}+1$. Each tetrahedron of the triangulation has six edges labeled by six spins $j_{e_1},..,j_{e_6}$ and we associate it with the corresponding \sj-symbol, which is the unique (non-trivial) $\SU(2)$ invariant built from these six representations. It is giving by combining four normalized Clebsh-Gordan coefficients corresponding to the four triangles of the tetrahedron. Finally, the Ponzano-Regge amplitude for a given colored triangulation is simply given by the product of the \sj-symbols associated to all its tetrahedra.

\begin{figure}[ht]
\begin{center}
\includegraphics[width=3cm]{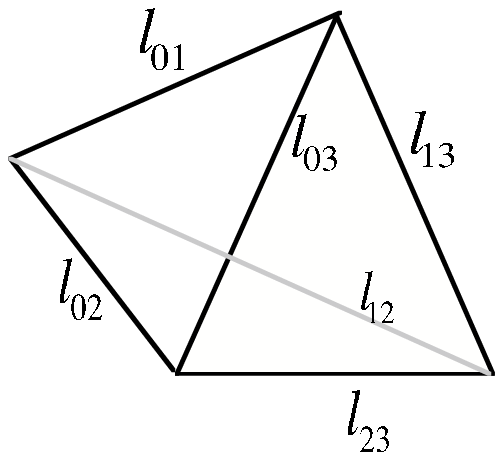}
\caption{A single tetrahedron: the edge lengths are given by $l_{IJ}=\frac{d_{j_{IJ}}}{2}$.}  \label{fig_tetrahedron}
\end{center}
\end{figure}

Looking more closely at a single tetrahedron, we label its four triangles by $I=0,..,3$. Then each of its six edges is labeled by the couple of triangles to which it belongs, $(IJ)$ with $0\leq I<J \leq 3$. To each edge is attached a $\SU(2)$ irrep of spin $j_{IJ}$, which defines the length of that edge $j_{IJ}+\frac{1}{2}=\frac{d_{j_{IJ}}}{2}$ (see figure \ref{fig_tetrahedron}).
There are several ways of expressing the \sj-symbol. The basic formula is the Racah's single sum formula which expresses the \sj-symbol as a sum over some products of factorials (see Appendix \ref{6j}). This is our starting point as in \cite{razvan}~:
\be
\label{Racah6j}
\left\{
\begin{array}{ccc}
j_{01}&j_{02}&j_{03}\\
j_{23}&j_{13}&j_{12}
\end{array}
\right\}
\,=
\sqrt{\Delta(j_{01}, j_{02}, j_{03}) \Delta(j_{23}, j_{02}, j_{12}) \Delta(j_{23}, j_{13}, j_{03}) \Delta(j_{01}, j_{13}, j_{12}) }
\sum_{\textrm{max } v_I}^{\textrm{min } p_J} \frac{(-1)^t (t+1)!}{\prod_I(t-v_I)! \prod_J (p_J-t)!}
\ee
where the $v_I$ and $p_i$ are given by the following sums:
$$
\forall K=0..3,\quad v_K= \displaystyle{\sum_{I \neq K}} j_{IK},\qquad
\forall k=1..3,\quad p_k= \displaystyle{\sum_{i \neq 0, k}} (j_{0i} + j_{ki}).
$$
The factors $\Delta(j_{01}, j_{02}, j_{03})$ are weights associated to each triangle and are defined by:
$$
\Delta(j_{01}, j_{02}, j_{03})= \frac{(j_{01}+j_{02}-j_{03})!(j_{01}-j_{02}+j_{03})!(-j_{01}+j_{02}+j_{03})!}{(j_{01}+j_{02}+j_{03}+1)!}.
$$
From this point, in all sums and products throughout this paper, capital indices $K$ will run from $0$ to $3$ and lower-cases indices $k$ will run from $1$ to $3$.

We are interested in the large spin expansion of the \sj-symbol when scaling all the spins homogeneously. Actually we will scale the lengths $d_{j_{IJ}}/2$ instead of the spins $j_{IJ}$ because the structure of the expansion will be simpler (we expect an alternation of cosines and sines without any mixing up at all orders as in \cite{valentin}) and the geometrical interpretation (when possible) is expected to be simpler. Then we rescale all $d_{j_{IJ}}$ by $\lambda d_{j_{IJ}}$ in (\ref{Racah6j}), which is equivalent to changing $j_{IJ}=\f{d_{j_{IJ}}}{2}-\f12$ to $\lambda\f{d_{j_{IJ}}}{2}-\f12$. This gives:
\equa{ \label{Racah_dj}
\tabl{l}{
\left\{ \tabl{lll}{\lambda d_{j_{01}}/2-1/2&\lambda d_{j_{02}}/2-1/2&\lambda d_{j_{03}}/2-1/2\\
\lambda d_{j_{23}}/2-1/2&\lambda d_{j_{13}}/2-1/2&\lambda d_{j_{12}}/2-1/2\\} \right\}= \\
\\
\;\;\;\;\;\sqrt{\Delta(\lambda d_{j_{01}}, \lambda d_{j_{02}}, \lambda d_{j_{03}}) \Delta(\lambda d_{j_{23}}, \lambda d_{j_{02}}, \lambda d_{j_{12}}) \Delta(\lambda d_{j_{23}}, \lambda d_{j_{13}}, \lambda d_{j_{03}}) \Delta(\lambda d_{j_{01}},\lambda d_{j_{13}}, \lambda d_{j_{12}}) }\\
\\
\;\;\;\;\; \displaystyle{\sum_{\lambda \,\textrm{max } \tilde{v}_I-\frac{3}{2}}^{\lambda \,\textrm{min } \tilde{p}_j-2}} (-1)^t \frac{(t+1)!}{\prod_i(t-\lambda \tilde{v}_I+\frac{3}{2})! \prod_j (\lambda \tilde{p}_j-t-2)!}
}}
with the new conventions:
$$
\tilde{v}_K= \sum_{I \neq K} \frac{d_{j_{IK}}}{2},\qquad
\tilde{p}_k= \sum_{i \neq 0, k} \frac{(d_{j_{0i}} + d_{j_{ki}})}{2},
$$
$$
\Delta(\lambda d_{j_{01}}, \lambda d_{j_{02}}, \lambda d_{j_{03}})= \frac{\left(\frac{\lambda }{2}(d_{j_{01}}+d_{j_{02}}-d_{j_{03}})-\frac{1}{2}\right)!\left(\frac{\lambda }{2}(d_{j_{01}}-d_{j_{02}}+d_{j_{03}})-
\frac{1}{2}\right)!\left(\frac{\lambda }{2}(-d_{j_{01}}+d_{j_{02}}+d_{j_{03}})-
\frac{1}{2}\right)!}{\left(\frac{\lambda }{2}(d_{j_{01}}+d_{j_{02}}+d_{j_{03}})-\frac{1}{2}\right)!}.
$$
The quantity $\tilde{v}_K$ gives the perimeter of the triangle $K$ while the $\tilde{p}_k$'s are the perimeters of (non-planar) quadrilaterals.

%%%%%%%%%%
\section{Perturbative expansion of the 6j-symbol}
%%%%%%%%%%%%
In this section, we will give a procedure to obtain the full perturbative expansion of the \sj-symbol in term of the length scale $\lambda$ and we compute explicitly the leading order (Ponzano-Regge formulae) then the next-to-leading order analytically.

%%%%%%
\subsection{General procedure}
%%%%%%
We give all the necessary formulae to obtain the Ponzano-Regge corrections at any order. But calculations are only performed explicitly at the next-to-leading order for a generic \sj-symbol. We start from equation (\ref{Racah_dj}).

%%%%%%%%%%%%
\paragraph{First approximation: factorials.}
%%%%%%%%%%%%

The factorial can be expanded in a series:
\equa{\label{fact}
 n!= \sqrt{2 \pi n}\left(\frac{n}{e}\right)^n\left(1 + \frac{1}{12n}+ \frac{1}{288n^2}- \frac{139}{51840n^3}-\frac{571}{2488320n^4} + \cdots \right)
}
In equation (\ref{Racah_dj}), there are factorials of the form: $n!$, $ (n+1/2)!$ and $(n-1/2)!$, which are rigourously defined through Euler's $\Gamma$ function. From (\ref{fact}) we deduce asymptotic expansions for $(n+1/2)!$ and $(n-1/2)!$ (see the details in appendix \ref{factorials}). In order to get the next-to-leading order (NLO) in the $1/\lambda$ expansion of the \sj-symbol, we replace the factorials in equation (\ref{Racah_dj}) by their respective asymptotic expansion:
\equa{ \tabl{l}{
n! \sim \sqrt{2 \pi }e^{(n+\frac{1}{2})\ln(n)-n}\left(1 + \frac{1}{12n}\right)\\
\\
(n+\frac{1}{2})! \sim \sqrt{2 \pi} e^{(n+1)\ln(n)-n}\left(1+ \frac{11}{24n}  \right) \\
\\
(n-\frac{1}{2})! \sim \sqrt{2 \pi} e^{n\ln(n)-n}\left(1- \frac{1}{24n}  \right). \\
}}
Then, equation (\ref{Racah_dj}) reads at first order as:
\equa{\label{6japprox1}
\left\{ \tabl{lll}{\lambda d_{j_{01}}/2-1/2&\lambda d_{j_{02}}/2-1/2&\lambda d_{j_{03}}/2-1/2\\
\lambda d_{j_{23}}/2-1/2&\lambda d_{j_{13}}/2-1/2&\lambda d_{j_{12}}/2-1/2\\} \right\}=
\frac{1}{2\pi} e^{\frac{\lambda }{2}h(d_{j_{IJ}})}
 \left(1 -\frac{1}{24\lambda } H(d_{j_{IJ}}) + O \left(\frac{1}{\lambda^2 }\right)\right) \,\Sigma\,.
}
The first factor is given by:
\bes
\label{termsh}
&& h(d_{j_{IJ}})= \sum_{I<J}d_{j_{IJ}}h_{d_{j_{IJ}}}\\
&&\textrm{with}\quad
h_{d_{j_{IJ}}}= \frac{1}{2} \ln \left(
\frac{(d_{j_{IJ}}-d_{j_{IK}}+d_{j_{IL}})(d_{j_{IJ}}+d_{j_{IK}}-d_{j_{IL}})(d_{j_{IJ}}-d_{j_{JK}}+d_{j_{JL}})
(d_{j_{IJ}}+d_{j_{JK}}-d_{j_{JL}})}{(d_{j_{IJ}}+d_{j_{IK}}+d_{j_{IL}})(-d_{j_{IJ}}+d_{j_{IK}}+d_{j_{IL}})
(d_{j_{IJ}}+d_{j_{JK}}+d_{j_{JL}})(-d_{j_{IJ}}+d_{j_{JK}}+d_{j_{JL}})}\right), \nonumber
\ees
where $(KL)$ is the opposite side to $(IJ)$, that is $K \neq L$ and $K,L \neq I,J$.
The second factor is due to the NLO of the factorials:
\be
\label{termsH}
H(d_{j_{IJ}})=2\displaystyle{\sum_{j,K}} \frac{1}{\tilde{p}_j-\tilde{v}_K} - 2\displaystyle{\sum_K} \frac{1}{\tilde{v}_K}= \displaystyle{\sum_I}\left[\frac{-r^I+\sum_{K \neq I}r^I_K}{2A_I}\right],
\ee
where $A_I$ is the area of triangle $I$, $r^I$ is the radius of the incircle of triangle $I$ and $r^I_K$  is the radius of the excircle to the triangle $I$ tangent to the side $d_{j_{IK}}$ of the triangle $I$.
Finally, $\Sigma$ is a Riemann sum:
\be
\label{termsSigma}
\Sigma=\frac{1}{\lambda ^2}\displaystyle{\sum_{x=\textrm{max } \tilde{v}_I/2}^{\textrm{min } \tilde{p}_j/2}} e^{F(x)} \left( 1 -\frac{1}{12\lambda }G(x) + O\left(\frac{1}{\lambda^2 }\right)\right) e^{\lambda f(x)}
\ee
with the pre-factor and the action given by:
\bes
f(x)&=& i \pi x+ x \ln(x) - \displaystyle{\sum_K}(x-\tilde{v}_K) \ln(x-\tilde{v}_K) - \displaystyle{\sum_j}(\tilde{p}_j-x) \ln(\tilde{p}_j-x), \nonumber\\
F(x)&=& \frac{1}{2} \ln \left( \frac{x^3 \prod_j(\tilde{p}_j-x)^3}{\prod_K (x-\tilde{v}_K)^4} \right),\\
G(x)&=&-\frac{13}{x}+\frac{47}{2} \displaystyle{\sum_K}\frac{1}{x-\tilde{v}_K} +13\sum_j \frac{1}{\tilde{p}_j-x}. \nonumber
\ees
The details of the computation are given in Appendix \ref{stirling}.

\medskip

%%%%
\paragraph{Second approximation: Riemann sum.}
%%%%

The second approximation consists in replacing the Riemann sum $\Sigma$ of (\ref{6japprox1}) by an integral. One $k^{-1}$ factor of $\Sigma$ plays the role of $dx$. We can then rewrite equation (\ref{6japprox1}) as:
\be
\{6j\}\sim\frac{1}{2\pi}  \left(1 -\frac{1}{24\lambda} H(d_{j_{IJ}}) + O \left(\frac{1}{\lambda}\right)\right) e^{\frac{\lambda}{2}h(d_{j_{IJ}})} \frac{1}{\lambda}\displaystyle{\int_{\textrm{max } \f{\tilde{v}_I}{2}}^{\textrm{min } \f{\tilde{p}_j}{2}}} dx \, e^{F(x)} \left( 1 -\frac{1}{12\lambda}G(x) + O\left(\frac{1}{\lambda^2}\right)\right) e^{\lambda f(x)}.
\ee
%
%\equa{
%\tabl{l}{
%\left\{ \tabl{lll}{kd_{j_{01}}/2-1/2&kd_{j_{02}}/2-1/2&kd_{j_{03}}/2-1/2\\
%kd_{j_{23}}/2-1/2&kd_{j_{13}}/2-1/2&kd_{j_{12}}/2-1/2\\} \right\}= \\
%\\
%\frac{1}{2\pi}  \left(1 -\frac{1}{24k} H(d_{j_{IJ}}) + O \left(\frac{1}{k}\right)\right) e^{\frac{k}{2}h(d_{j_{IJ}})} \frac{1}{k}\displaystyle{\int_{\textrm{max } \tilde{v}_I/2}^{\textrm{min } \tilde{p}_j/2}} dx e^{F(x)} \left( 1 -\frac{1}{12k}G(x) + O\left(\frac{1}{k}\right)\right) e^{kf(x)}
%}}
This approximation does not generate any corrections at leading order and at first order. It will nevertheless enter at second order in terms in $1/\lambda^2$.
%In Appendix \ref{riemann}, we show that this approximation is $O(1/k^2)$.

\paragraph{Third approximation: saddle point approximation.}
We have to study an integral of the form $I=\int_a^b dx g(x) e^{\lambda f(x)}$ where $\lambda$ is a large parameter. The asymptotic expansion of such an integral is given by contributions around the stationary points of the action $f$ which are  points, denoted $x_0$, of the complex plane such that $f^{\prime}(x_0)=0$. We expand the action $f(x)$ and the function $g(x)$ around the stationary points $x_0$ in term of $\delta x=x-x_0$:
$$
f(x)=\displaystyle{\sum_{j=0}^\infty}\frac{f(x_0)^{(j)}}{j!}(\delta x)^j=f(x_0)+\frac{f^{\prime \prime }(x_0)}{2}(\delta x)^2+ f_{x_0}^{>2}(\delta x)  \quad \textrm{and}\quad g(x)=\displaystyle{\sum_{j=0}^\infty}\frac{g(x_0)^{(j)}}{j!}(\delta x)^j=g(\delta x).
$$
We then expand $K(\delta x)=g(\delta x)e^{kf_{x_0}^{>2}(\delta x)}$ in power of $\delta x$. Following the standard stationary phase approximation, we extend the integration domain to the whole $\R$.  The integrals are then ``generalized Gaussians" which can easily be computed. We  group the resulting terms according to their dependence on $1/\lambda$, being careful because of the function $g(x)$ which  depends on $1/\lambda$. We recall that $g(x)$  was obtained by replacing the factorials in (\ref{Racah_dj}) by their series expansion and we write $g(x)$ under the general form:
$$
g(x)=\displaystyle{\sum_{i=1}^\infty}\frac{g_i(x)}{i!\,\lambda^i}.
 $$
Then the complete perturbative expansion of $I$ can be written as:
\equa{\label{complete}
I= \displaystyle{\sum_{x_0}} e^{\lambda f(x_0)}\sqrt{\frac{2\pi}{-f^{\prime \prime}(x_0)\lambda}} \left(1+\displaystyle{\sum_{n=1}^\infty}\frac{1}{\lambda^n} \left[ \displaystyle{\sum_{p=0}^{n-1}} \tilde{N}_p\frac{(2p-1)!!}{(-f^{\prime \prime}(x_0))^p} + \displaystyle{\sum_{p=0}^{2n}}N_p \frac{(2n+2p-1)!!}{(-f^{\prime \prime}(x_0))^{n+p}} \right] \right)
}
where \equa{ \tabl{l}{
\tilde{N}_p=\displaystyle{\sum_{i=1}^{E[\frac{2p}{3}]}} \frac{1}{i!(n-p+i)!} \displaystyle{\sum_{l_1 \cdots l_i=3}^{E[\frac{2p}{i}]}} \frac{g_{n-p+i}^{(2p-\sum_{j=1}^il_j)}(x_0)}{(2p-\sum_{j=1}^il_j)!} \displaystyle{\prod_{j=1}^i} \frac{f^{l_j}(x_0)}{(l_j)!}\\
\\
N_0=\frac{g_0^{(2n)}(x_0)}{(2n)!}+\displaystyle{\sum_{i=1}^{E[\frac{2n}{3}]}} \frac{1}{(i!)^2} \displaystyle{\sum_{l_1 \cdots l_i=3}^{E[\frac{2n}{i}]}} \frac{g_{i}^{(2n-\sum_{j=1}^il_j)}(x_0)}{(2p-\sum_{j=1}^il_j)!} \displaystyle{\prod_{j=1}^i} \frac{f^{l_j}(x_0)}{(l_j)!}\\
\\
N_p=\displaystyle{\sum_{i=p}^{E[\frac{2(p+n)}{3}]}} \frac{1}{i!(i-p)!} \displaystyle{\sum_{l_1 \cdots l_i=3}^{E[\frac{2(n+p)}{i}]}} \frac{g_{i-p}^{(2(n+p)-\sum_{j=1}^il_j)}(x_0)}{(2(n+p)-\sum_{j=1}^il_j)!} \displaystyle{\prod_{j=1}^i} \frac{f^{l_j}(x_0)}{(l_j)!} \textrm{ for } p\geq1\\
}}
The details of the computation are given in Appendix \ref{saddlepoint}. From this expansion and adjusting the first approximation
%, where we replaced the factorials by their asymptotic expansion,
to get the proper dependence on $\lambda$ for $g$ and the pre-factors, it is possible to compute analytically the whole asymptotic expansion of the \sj-symbol.

Here to get  explicitly the next-to-leading order of the \sj-symbol asymptotic expansion, we only  need the next-to-leading order of the $1/\lambda$ expansion of $I$, so we cut the previous formulae at $n=1$, then
$$
I \sim  \displaystyle{\sum_{x_0}} e^{\lambda f(x_0)}\sqrt{\frac{2\pi}{-f^{\prime \prime}(x_0)\lambda }} \left(1+\frac{1}{\lambda }\left(\tilde{N}_0+ \frac{N_0}{-f^{\prime \prime }(x_0)}+\frac{3N_1}{(-f^{\prime \prime }(x_0))^2}+\frac{15N_2}{(-f^{\prime \prime }(x_0))^3}\right) \right)
$$
with the expansion coefficients given by
$$
\tilde{N}_0=g_1(x_0), \quad
N_0=\frac{g_0^{\prime \prime }(x_0)}{2}, \quad N_1=\frac{f^{(3)}(x_0)g_0^\prime(x_0)}{3!}+\frac{f^{(4)}(x_0)g_0(x_0)}{4!}, \quad N_2=\frac{g_0(x_0)}{2}\left(\frac{f^{(3)}(x_0)}{3!}\right)^2.
$$
We recall that $g(x)=e^{F(x)}\left(1-\frac{G(x)}{12\lambda }\right)$; that is: $g_0(x)=e^{F(x)}$ and $g_1(x)=-\frac{G(x)}{12}e^{F(x)}.$  Finally, we obtain the approximation:
\equa{\tabl{ll}{\label{intNLO}
I \sim \displaystyle{\sum_{x_0}} &\sqrt{\frac{2\pi}{-f^{\prime \prime}(x_0)\lambda} }\, e^{F(x_0)+\lambda f(x_0)} \\
&\left[1 +
\frac{1}{\lambda } \left(- \frac{G(x_0)}{12}- \frac{F^{\prime \prime}(x_0)+(F^{\prime}(x_0))^2}{2 f^{\prime \prime}(x_0)}+\frac{f^{(4)}(x_0)+4f^{(3)}(x_0)F^\prime(x_0)}{8(f^{\prime \prime}(x_0))^2}- \frac{5(f^{(3)}(x_0))^2}{24(f^{\prime \prime}(x_0))^3} \right) + O \left(\frac{1}{\lambda^2} \right) \right]
}}
This gives us the following expression for the asymptotic expansion of the \sj-symbol at second order:
\equa{\label{NLO1}
\tabl{l}{
\left\{ \tabl{lll}{\lambda d_{j_{01}}/2-1/2&\lambda d_{j_{02}}/2-1/2&\lambda d_{j_{03}}/2-1/2\\
\lambda d_{j_{23}}/2-1/2&\lambda d_{j_{13}}/2-1/2&\lambda d_{j_{12}}/2-1/2\\} \right\} \\
\\
\;\;\;\;\;\;\;\; \sim \displaystyle{\sum_{x_0}} \sqrt{\frac{1}{-f^{\prime \prime}(x_0)2\pi \lambda^{3}} } \exp\left(F(x_0)+\lambda f(x_0)\right) \exp\left(\displaystyle{\sum_{I<J}}\frac{\lambda d_{j_{IJ}}}{2}h_{d_{j_{IJ}}}\right) \\
\\
\;\;\;\;\;\;\;\;\;\;\; \left[1 +
\frac{1}{\lambda} \left(- \frac{H(j_{IJ})}{24}-\frac{G(x_0)}{12}- \frac{F^{\prime \prime}(x_0)+(F^{\prime}(x_0))^2}{2 f^{\prime \prime}(x_0)}+\frac{f^{(4)}(x_0)+4f^{(3)}(x_0)F^\prime(x_0)}{8(f^{\prime \prime}(x_0))^2}- \frac{5(f^{(3)}(x_0))^2}{24(f^{\prime \prime}(x_0))^3} \right) + O \left(\frac{1}{\lambda^2} \right) \right]
}}
where $x_0$ are the stationary points of the phase, i.e. $f^\prime(x_0)=0$. The next step is to identify these stationary points.

%%%%%%
\subsection{Contributions of the stationary points}
%%%%%%

The phase $f(x)$ is an analytical function given by:
\equa{
f(x)= i \pi x+ x \ln(x) - \displaystyle{\sum_K}\left(x-\frac{\tilde{v}_K}{2}\right) \ln\left(x-\frac{\tilde{v}_K}{2}\right) - \displaystyle{\sum_j}\left(\frac{\tilde{p}_j}{2}-x\right) \ln\left(\frac{\tilde{p}_j}{2}-x\right)
}
therefore the stationary points $x_0$ satisfy the following equation as shown in \cite{razvan}:
 \equa{ \label{stationaryequa}
  f^\prime(x)= i\pi + \ln(x)-\sum \ln\left(x-\tilde{v}_K/2\right) + \sum \ln\left(\tilde{p}_j/2-x\right)=0
  }
  which is equivalent to
  \equa{\label{stationaryequa2}
  x \prod_j (p_j-x)=-\prod_K(x-v_K)
  }
The previous equation reduces to a quadratic equation $Ax^2-Bx+C=0$ with
 \equa{
 \tabl{l}{
A=-\displaystyle{\sum_{j<l}}\tilde{p}_k \tilde{p}_l+\displaystyle{\sum_{K<L}}\tilde{v}_K \tilde{v}_L=\frac{1}{2}\left(\displaystyle{\sum_{\tabl{c}{I<J, K<L,\\ (I,J) \neq (K,L)}}d_{j_{IJ}}d_{j_{KL}}}\right)\\
B=-\tilde{p}_1\tilde{p}_2\tilde{p}_3+\displaystyle{\sum_{I<J<K}}\tilde{v}_I\tilde{v}_J\tilde{v}_K=\frac{1}{4}\left[\left(\displaystyle{\sum_{\tabl{c}{I<J, K<L,\\ (I,J) \neq (K,L)}}d_{j_{IJ}}d_{j_{KL}}}\right)\left(\displaystyle{\sum_{I<J}}d_{j_{IJ}}\right)+\displaystyle{\sum_J}\left(\displaystyle{\prod_{K \neq J}}d_{j_{JK}}\right)\right]\\
C=\displaystyle{\prod_K}v_K\\
}}
As shown in \cite{razvan}, the discriminant $\Delta= -(B^2-4AC)$ is given in terms of the $d_{j_{IJ}}$ by:
\equa{\tabl{ll}{
\Delta&=\frac{1}{16}\left[ \displaystyle{\sum_{\tabl{c}{I<J,\\ K<L,\\ (I,J)\neq (K,L)}}}d_{j_{IJ}}d_{j_{KL}}\left(\displaystyle{\sum_{\tabl{c}{M<N,\\ (M,N)\neq (I,J),\\(M,N)\neq (K,L)}}}d_{j_{MN}}^2-d_{j_{IJ}}^2-d_{j_{KL}}^2\right)- \displaystyle{\sum_K}\displaystyle{\prod_{L\neq K}}d_{j_{KL}}^2\right] \\
&\\
&=2 \begin{vmatrix}
0 &\left(\frac{d_{j_{23}}}{2}\right)^2 & \left(\frac{d_{j_{13}}}{2}\right)^2&\left(\frac{d_{j_{12}}}{2}\right)^2& 1 \\
\left(\frac{d_{j_{23}}}{2}\right)^2 &0&\left(\frac{d_{j_{03}}}{2}\right)^2&\left(\frac{d_{j_{02}}}{2}\right)^2 & 1 \\
\left(\frac{d_{j_{13}}}{2}\right)^2 &\left(\frac{d_{j_{03}}}{2}\right)^2  & 0&\left(\frac{d_{j_{01}}}{2}\right)^2&1 \\
\left(\frac{d_{j_{12}}}{2}\right)^2 & \left(\frac{d_{j_{02}}}{2}\right)^2 &\left(\frac{d_{j_{01}}}{2}\right)^2&0&1\\
1&1&1&1&0\\
\end{vmatrix} =2^4(3!)^2V^2
}}
where $V$ is the volume of the tetrahedron of edge length $d_{j_{IJ}}/2$. In the following we will focus on the case where $\Delta>0$, i.e. $V^2>0$, which corresponds to tetrahedra in flat Euclidean space. The other case $\Delta<0$ corresponds to tetrahedra admitting an embedding in the 2+1d Minkowski space-time. And so, we get two stationary points:
\equa{ \label{statiopoints}
x_\pm=\frac{B\pm i \sqrt{\Delta}}{2A}
}
The geometrical interpretation of the stationary points is not clear yet. We have shown that $\Delta$ is related to the volume of the tetrahedron. $B$ and $A$ are also related to invariant of the tetrahedron:
$$B=\sum_I \frac{v_I}{2} A + 24V \cot \theta$$
where we recall that $v_I$ is the perimeter of the triangle I of the tetrahedron. The angle $\theta$ is the Brocard angle of the tetrahedron. Indeed, $\frac{d_{j_{01}}}{2} \frac{d_{j_{02}}}{2}\frac{d_{j_{03}}}{2} ::\frac{d_{j_{03}}}{2} \frac{d_{j_{23}}}{2}\frac{d_{j_{13}}}{2} :: \frac{d_{j_{12}}}{2} \frac{d_{j_{02}}}{2}\frac{d_{j_{23}}}{2}  :: \frac{d_{j_{01}}}{2} \frac{d_{j_{12}}}{2}\frac{d_{j_{13}}}{2} $ are the barycentric coordinates of the second Lemoine point of the tetrahedron denoted $L$. This point is such that the distance from $L$ to the face $I$ of the tetrahedron is equal to $R_I \tan \theta$ where $R_I$ is the radius of the circumscribed circle of the triangle $I$ and $\theta$ is then defined by $\displaystyle{\sum_J}\left(\displaystyle{\prod_{K \neq J}}\frac{d_{j_{JK}}}{2}\right)= 12V \cot \theta$.

The geometrical significance of the stationary points still has to be understood. However, we can now give the explicit form of the leading order and of the next to leading order of the \sj-symbol.

\medskip

\textit{Leading order.}
We first focus on the leading order and on the $x_+$ contribution. This analysis has already been done in \cite{razvan} and we just recall the main steps and give the notations:

 $f(x_+)= \displaystyle{\sum_{I<J}}\frac{d_{j_{IJ}}}{2}f_{d_{j_{IJ}}}$ where
\equa{ \label{functionf}
\tabl{l}{
f_{d_{j_{0i}}}= \ln\left[\frac{(x_+-\tilde{v}_0)(x_+-\tilde{v}_i)}{\prod_{j\neq i}(\tilde{p}_j-x_+)}\right] \textrm{ for } i, j \in \{1, \cdots, 3\} \\
\\
f_{d_{j_{ik}}}=\ln\left[\frac{(x_+-\tilde{v}_k)(x_+-\tilde{v}_i)}{(\tilde{p}_k-x_+)(\tilde{p}_i-x_+)}\right] \textrm{ for } i, k \in \{1, \cdots, 3\} \\
}}

The second derivative of $f$ is given by:
$$\tabl{ll}{-f^{\prime \prime}(x_+)&= \sum_K \frac{1}{x_+-\tilde{v}_K}+\sum_j\frac{1}{\tilde{p}_j-x_+}-\frac{1}{x_+}\\
&\\
&=-i \sqrt{\Delta}\exp(-\ln(x_+\prod_j(\tilde{p}_j-x_+)))
}$$
where we have used the equation (\ref{stationaryequa2}) which gives $x_+\prod_j (\tilde{p}_j-x_+)=-\prod_K(x_+-\tilde{v}_K)$. In the same way, we can simplify $F(x_+)=-\frac{1}{2} \ln\left( x_+ \prod_j \left( \tilde{p}_j-x_+ \right) \right)$. The exponential piece of $f^{\prime \prime}(x_+)$ and $e^{F(x_+)}$ compensate and we get:
$$ \frac{1}{\sqrt{-f^{\prime \prime}(x_+)}} e^{F(x_+)}= \frac{1}{\sqrt{-i\sqrt{\Delta}}}$$

Collecting these different results yields  the following contribution of the $x_+$ stationary point:
\equa{\tabl{l}{
 \sqrt{\frac{1}{-f^{\prime \prime}(x_+)2\pi \lambda^{3}} } \exp\left(F(x_+)+\lambda f(x_+)\right) \exp\left(\displaystyle{\sum_{I<J}}\frac{\lambda d_{j_{IJ}}}{2}h_{d_{j_{IJ}}}\right)=\\
 \;\;\;\;\;\frac{1}{\sqrt{2\pi \lambda^3\sqrt{\Delta}}} \exp\left[ i \frac{\pi}{4}+ \sum_{IJ}(\lambda d_{j_{IJ}}/2)(h_{d_{j_{IJ}}}+f_{d_{j_{IJ}}})\right]
}}

The same analysis for the $x_-$ contribution yields the same contribution as the previous one with an opposite phase:
\equa{\tabl{l}{
 \sqrt{\frac{1}{-f^{\prime \prime}(x_-)2\pi \lambda^{3}} } \exp\left(F(x_-)+\lambda f(x_-)\right) \exp\left(\displaystyle{\sum_{I<J}}\frac{\lambda d_{j_{IJ}}}{2}h_{d_{j_{IJ}}}\right)=\\
 \;\;\;\;\;\frac{1}{\sqrt{2\pi \lambda^3\sqrt{\Delta}}} \exp\left[ -i \frac{\pi}{4}+ \sum_{IJ}(\lambda d_{j_{IJ}}/2)(h_{d_{j_{IJ}}}+\overline{f_{d_{j_{IJ}}}})\right]}}

We must now compute $f_{d_{j_{IJ}}}$ which is a complex logarithm. We recall that the principal value of the logarithm is defined by $\textrm{Log} z := \ln |z| +i \textrm{Arg} z$. Therefore, we have to compute $\Im(f_{d_{j_{IJ}}})=\theta_{IJ}$.  From (\ref{functionf}), we can write that:
\equa{ \label{angle}
\tabl{l}{
\theta_{0i}= \textrm{Arg} (x_+-\tilde{v}_0) +\textrm{Arg} (x_+-\tilde{v}_i) -  \displaystyle{\sum_{j \neq i}}\textrm{Arg} (\tilde{p}_j-x_+) \\
\theta_{ik}= \textrm{Arg} (x_+-\tilde{v}_k) +\textrm{Arg} (x_+-\tilde{v}_i) -  \textrm{Arg} (\tilde{p}_k-x_+)- \textrm{Arg} (\tilde{p}_i-x_+) \\
}}
The analysis done in \cite{razvan} shows that $\theta_{IJ}$ can be identified as the (exterior) dihedral angles of the tetrahedron. Moreover,
\equa{\tabl{l}{
\Re(f_{d_{j_{0i}}})=\ln \left| \frac{(x_+-\tilde{v}_0)(x_+-\tilde{x}_i)}{\prod_{j\neq i} (\tilde{p}_j-x_+)} \right|\\
\\
\Re(f_{d_{j_{ik}}})=\ln \left| \frac{(x_+-\tilde{v}_k)(x_+-\tilde{v}_i)}{(\tilde{p}_i-x_+)(\tilde{p}_k-x_+)} \right|
}}
A tedious (but interesting) computation shows that:
\equa{
\Re(f_{d_{j_{IJ}}}) + h_{d_{j_{IJ}}}=0
}
Then, summing the contributions of $x_+$ and $x_-$ we get the leading order of the 6j-symbol:
\equa{\label{LO}
\left\{ \tabl{lll}{\lambda d_{j_{01}}/2-1/2&\lambda d_{j_{02}}/2-1/2&\lambda d_{j_{03}}/2-1/2\\
\lambda d_{j_{23}}/2-1/2&\lambda d_{j_{13}}/2-1/2&\lambda d_{j_{12}}/2-1/2\\} \right\}
 \stackrel{\textrm{L.O.}}{\sim}\sqrt{ \frac{1}{12\pi \lambda^3 V}} \cos\left[  \frac{\pi}{4}+S_R \right]
}
where $S_R= \sum_{I<J}\frac{\lambda d_{j_{IJ}}}{2}\theta_{IJ}$ is the Regge action. This is the well-known limit given by Ponzano and Regge \cite{PR} and which has justified their state sum model for 3d Euclidean gravity where the \sj-symbol is the spinfoam amplitude for a single tetrahedron. %Indeed, Regge calculus is a discrete approximation to general relativity.

\medskip

\begin{figure}[ht]
\includegraphics[width=5cm]{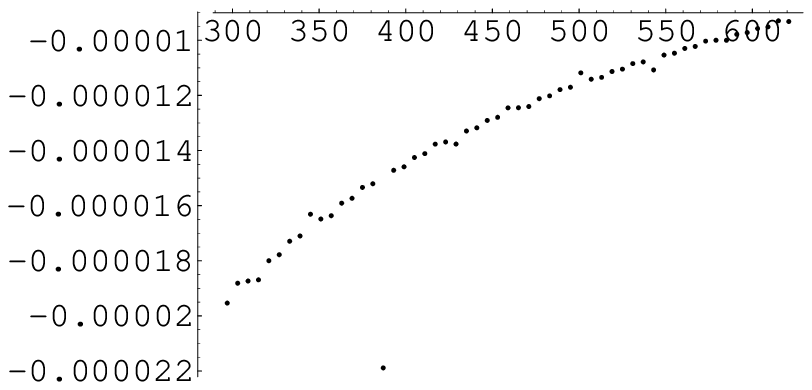} \;\;\;\;\;\;\;\;\;\;\;\;\;\;\;\;\;\;\;\;\;\;
\includegraphics[width=5cm]{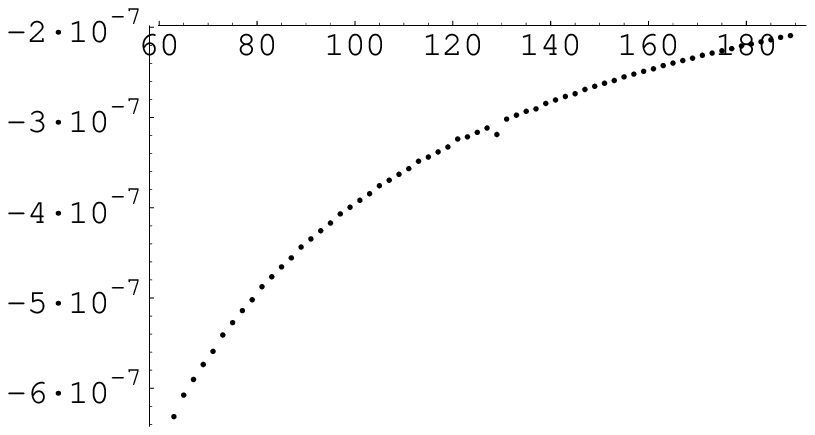}
\caption{Plots of the difference $\delta_{NLO}$ between the \sj-symbol and its analytical approximation up to NLO. On the left, we look at the \sj-symbol for $d_1=5\lambda, d_2=7\lambda, d_3=9\lambda, d_4=7\lambda, d_5=9\lambda, d_6=9\lambda$ with the x-coordinate standing for $3\lambda$. On the right, we've plotted the case $d_1=15\lambda, d_2=17\lambda, d_3=19\lambda,d_4=19\lambda, d_5=21\lambda, d_6=17\lambda$ with $\lambda$ running from 60 to 200.} \label{geneplot}
\end{figure}

\textit{Next to leading order.}
The next-to-leading order is then given by the term in $\frac{1}{\lambda^{5/2}}$ in equation (\ref{NLO1}). Using equations (\ref{termsH}-\ref{termsh}-\ref{termsSigma}), we rewrite the leading order in terms of $x_\pm, \tilde{v}_I, \tilde{p}_j$ and $\Delta$:
\equa{\label{NLO}
\frac{1}{\sqrt{48\pi \lambda^5 V}}\left\{ A(x_+, \tilde{v}_I, \tilde{p}_j,\Delta) e^{i(S_R+\frac{\pi}{4})}+A(x_-, \tilde{v}_I, \tilde{p}_j,\Delta) e^{-i(S_R+\frac{\pi}{4})}\right\}
}
where
\equa{\tabl{l}{
A(x_+, \tilde{v}_I, \tilde{p}_j,\Delta)=- \frac{H(d_{j_{IJ}})}{24}+\frac{1}{24i \sqrt{\Delta}\Delta \prod_I (x_+-\tilde{v}_I)} [-\Delta^2-3i (\displaystyle{\sum_K} \displaystyle{\prod_{L\neq K}} (x_+-\tilde{v}_L) ) \Delta \sqrt{\Delta}
\\ \;\;\; \;\;\;\;\;\;+( 9\displaystyle{\sum_K} \displaystyle{\prod_{L\neq K}} (x_+-\tilde{v}_L)^2 +6 \displaystyle{\prod_I} (x_+-\tilde{v}_I) \displaystyle{\sum_{K<L}} (x_+-\tilde{v}_K)(x_+-\tilde{v}_L))\Delta\\
\;\;\;\;\;\;\;\;\; -6i( -\displaystyle{\prod_j}(\tilde{p}_j -x_+)^3-\displaystyle{\sum_K}\displaystyle{\prod_{L\neq K}}(x_+-\tilde{v}_L)^3 +\displaystyle{\sum_j} x_+^3\displaystyle{ \prod_{l\neq j}} (\tilde{p}_l-x_+)^3\\
\;\;\;\;\;\;\;\;\;\;\;\;\;\;\;\;\;\;-(\displaystyle{\sum_K} \displaystyle{\prod_{L\neq K}} (x_+-\tilde{v}_L))(-\displaystyle{\prod_j}(\tilde{p}_j -x_+)^2+\displaystyle{\sum_K}\displaystyle{\prod_{L\neq K}}(x_+-\tilde{v}_L)^2 -\displaystyle{\sum_j} x_+^2\displaystyle{ \prod_{l\neq j}} (\tilde{p}_l-x_+)^2) )\sqrt{\Delta}\\
\;\;\;\;\;\;\;\;\; -5(-\displaystyle{\prod_j}(\tilde{p}_j -x_+)^2+\displaystyle{\sum_K}\displaystyle{\prod_{L\neq K}}(x_+-\tilde{v}_L)^2 -\displaystyle{\sum_j} x_+^2\displaystyle{ \prod_{l\neq j}} (\tilde{p}_l-x_+)^2)^2  ]
}}
Since $x_\pm$ are conjugated to each other, we obviously have $A(x_+)=\overline{A(x_-)}$.
Moreover, numerical computations shows that $\Re(A(x_\pm, \tilde{v}_I, \tilde{p}_j,\Delta))=0$, and in particular $A(x_+)=-{A(x_-)}$. This is a priori a non-trivial result to obtain from the previous formulas. Nevertheless, we tested it numerically for various choices of spin and it always turned out true. Thus we believe that there should be a way to show it analytically.
%$A(x_+, \tilde{v}_I, \tilde{p}_j,\Delta)=-A(x_-, \tilde{v}_I, \tilde{p}_j,\Delta)$.
We can then give an explicit formula of the NLO of the \sj-symbol:
\be
\label{NLOsin}
\{6j\}\underset{\lambda\arr\infty}{\sim}
\{6j\}_{NLO}=
\frac{1}{\sqrt{ 12\pi \lambda^3 V}} \cos\left[  \frac{\pi}{4}+S_R \right]
-\frac{1}{\sqrt{12\pi \lambda^5 V}}\Im(A(x_+,\tilde{v}_I, \tilde{p}_j,\Delta)) \sin(S_R+\pi/4).
\ee
This result is confirmed by numerical simulations. The plots figure \ref{geneplot} represent numerical simulations of the \sj-symbol minus its approximation given above (\ref{NLOsin}). Moreover, to enhance the comparison, we have multiplied by $\lambda^{5/2}$ to see how the coefficient of the next to leading order is approached and we have divided by $\cos (S_R+\pi/4)$ (oscillations of the next-to-next-to-leading order) to suppress the oscillations; that is we have plotted:
\be
\delta_{NLO}\,\equiv\,\lambda^{5/2}\,\f{\{6j\}-\{6j\}_{NLO}}{\cos (S_R+\pi/4)}.
\ee
As expected, the numerical simulations show that this rescaled difference $\delta_{NLO}$ goes to 0 as $1/\lambda$ when $\lambda$ goes to $\infty$. Moreover, the data for $\delta_{NLO}$ without any oscillation suggest that we correctly divided by $\cos (S_R+\pi/4)$ and thus the NNLO of the\sj-symbol should oscillate in $\cos (S_R+\pi/4)$. Therefore, this strongly suggest that the asymptotic expansion of the \sj-symbol in term of the length scale $\lambda$ is given by an alternative of cosines and sinus at each order. We strongly underline that this is true because we have rescaled the edge lengths $d_{j_{IJ}}$. If we had instead rescaled the spins $j_{IJ}$ as usually done, we would have found an oscillatory behavior controlled by a mixing of $\cos$ and $\sin$ at each order (as shown explicitly for the case of the isosceles tetrahedron in \cite{valentin}). This suggests that the $d_{j_{IJ}}$ are indeed the right parameter to consider when studying the semi-classical behavior of the \sj-symbol.

The only thing left to do in the present analysis is to provide the NLO coefficient $\Im(A(x_+))$ with a geometrical interpretation and to show rigourously that $\Re(A(x_+))$ vanishes.

Finally, we rewrite the approximation up to NLO of the \sj-symbol in a slightly different manner:
\be
\{6j\}\sim
\frac{1}{\sqrt{ 12\pi \lambda^3 V}} \cos\left[  \frac{\pi}{4}+S_R + \f1\lambda\Im(A(x_+)) +O\left(\f1{\lambda^2}\right)\right].
\ee
This shows that the next-to-leading corrections to the \sj-symbol can be directly considered as corrections to the Regge action for (3d) gravity:
$$
S_R^{corrected}\,\equiv\,S_R + \f1\lambda\Im(A(x_+)).
$$
We point out that an expansion in $1/\lambda$ with alternating $\cos$ and $\sin$ could be similarly re-absorbed as corrections to the Regge action. This would define in the spinfoam framework the quantum gravity corrections to classical 3d gravity due to the fundamental discreteness of the theory. Such correction would enter the gravitational correlations (of the ``graviton propagator" type) at second order as suggested in  \cite{josh}.

%This supports the ``graviton" calculations to done for the isosceles tetrahedron \cite{josh}.

%%%%%%%%%%%%%%%%%%%%%%%%%
%\subsection{Geometrical interpretation}
%%%%%%%%%%
%%%%%%%%%%%%%%%
\section{Some particular cases}
%%%%%%%%%%%%%%%

%%%%%%%%
\subsection{The equilateral tetrahedron}
%%%%%%%%

\begin{figure}[ht]
\begin{center}
\includegraphics[width=5cm]{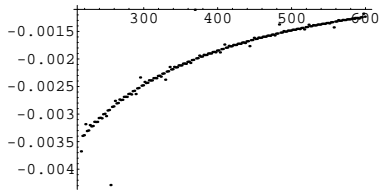}
\caption{Difference between the equilateral \sj-symbol and the analytical result (\ref{NLOequa}). The x-axis stands for $d$ and $d$ goes from 200 to 600.} \label{equiplot}
\end{center}
\end{figure}

For the equilateral tetrahedron, all the edges have the same length: that is $\forall I,J,\,\, d_{j_{IJ}}=d$. The tetrahedron with edge length $d/2$ has a volume $V=(d/2)^3\,\sqrt{2}/12$ and has all equal dihedral angles $\theta=\arccos(-1/3)$.
In this case, the expressions greatly simplify. For instance, the stationary points are $x_\pm= \frac{11 \pm i\sqrt{\frac{1}{2}}}{6}d$. Equations (\ref{LO}) and (\ref{NLO})reduce to:
\equa{\label{NLOequa}
\{6j\}^{\textrm{NLO}}_{\textrm{equi}} =\frac{2^{5/4}}{\sqrt{\pi d^3}} \cos\left(S_R+\frac{\pi}{4}\right)-\frac{31}{72\sqrt{\sqrt{2}\pi d^5}}\sin\left(S_R+\frac{\pi}{4}\right)
}
where the Regge action is $S_R=3d\theta$. The result was already obtained in \cite{valentin}. We confirm it by numerical simulations. The plot in fig.\ref{equiplot} gives the equilateral \sj-symbol minus its NLO approximation (\ref{NLOequa}). Like for the previous plots, we have multiplied by $\lambda^{5/2}$ to see how the coefficient of the next to leading order is approached and we have divided by $\cos (S_R+\pi/4)$ (oscillations of the next to next to leading order) to suppress the oscillations.

%%%%%%%
\subsection{The isosceles thetrahedron}
%%%%%%%

\begin{figure}[ht]
\begin{center}
\includegraphics[width=3cm]{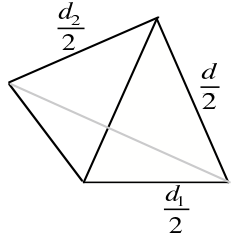}
\caption{The isosceles tetrahedron} \label{fig_isotetra}
\end{center}
\end{figure}

We now consider an isosceles tetrahedron that is a tetrahedron which has two opposite edges of length equal to $\frac{d_1}{2}$ and $\frac{d_2}{2}$ and the remaining four edges of the same length equal to $\frac{d}{2}$ (see figure \ref{fig_isotetra}). The volume of the tetrahedron is:
$$
V^2=\frac{1}{2^8 (3!)^2} d_1^2d_2^2\left( 4{d}^{2}-d_1^{2}-d_2^{2} \right),
$$
and the dihedral angles are:
$$
\theta= \arccos \left( \frac{-d_1d_2}{\sqrt{4d^2-d_1^2}\sqrt{4d^2-d_2^2}}\right),\qquad \theta_{1,2}=2\arccos \left( \frac{d_{2,1}}{\sqrt{4d^2-d_{1,2}^2}}\right).
$$
Once again, equations (\ref{LO}) and (\ref{NLO}) simplify and we get~:
\equa{ \label{isoequa}
\{6j\}_{\textrm{NLO}}^{(\textrm{iso})}= \frac{1}{\sqrt{12\pi V\lambda^3}}\, \cos \left(S_R +\frac{\pi}{4}\right) -  \frac{F(d,d_1,d_2)}{24V \lambda\sqrt{12\pi V\lambda^3}}\,  \sin \left(S_R +\frac{\pi}{4}\right)
}
where
$
F(d,d_1,d_2)=\frac {768{d}^{6}(d^2-d_1^{2}-d_2^{2})+736{d}^{4}d_1^{2}d_2^{2}
+240d^{4}(d_1^{4}+d_2^{4})-176{d}^{2}d_1^2d_2^{2}(d_1^2+d_2^2)-24{d}^{2}(d_1^{6}+d_2^{6})+10d_1^2 d_2^{2}(d_1^4+d_2^4)+25 d_1^{4}d_2^{4}}
{96\left( 4d^2-d_1^2 \right)\left(4d^2-d_2^2 \right)\left( 4{d}^{2}-d_1^{
2}-d_2^{2} \right)},
$
and the Regge action $S_R=2d \theta+\frac{d_1}{2}\theta_1+\frac{d_2}{2}\theta_2$.  Let us point out that the volume increases as $\lambda^3$ while $F$ goes as $\lambda^2$, so that the NLO scales properly as $\lambda^{-5/2}$.

\begin{figure}[ht]
\includegraphics[width=5cm]{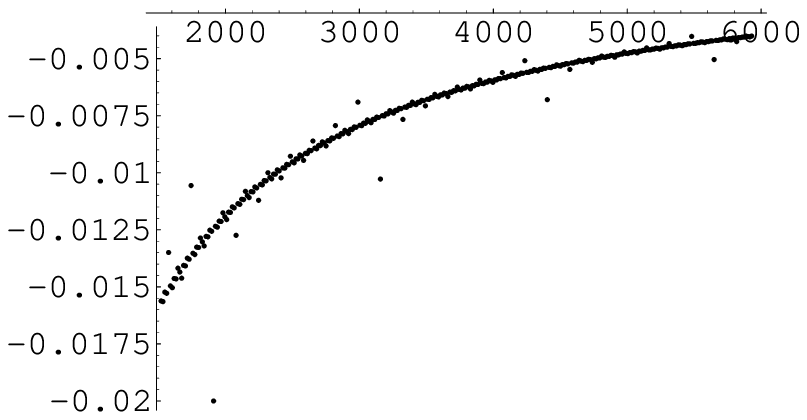} \;\;\;\;\;\;\;\;\;\;\;\;\;\;
\includegraphics[width=5cm]{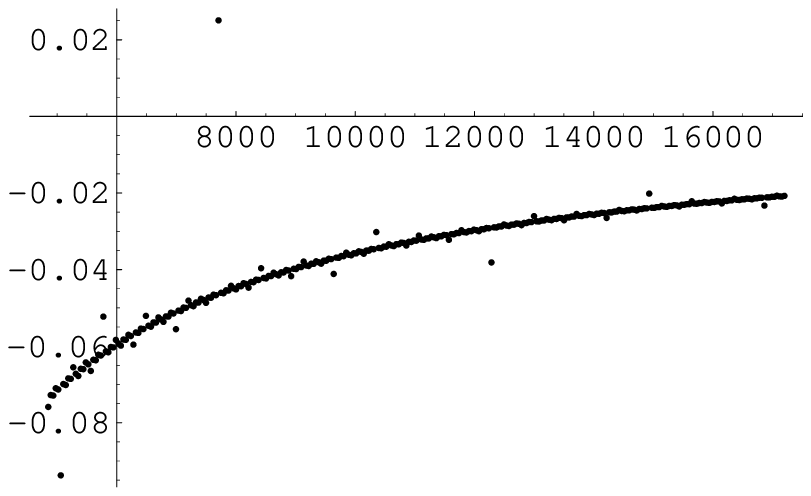}
\caption{Differences between isosceles \sj-symbols and their analytical approximation (\ref{isoequa}). The x-axis stands for $d$ with $\lambda$ goes from 200 to 600. On the left hand side, we consider isosceles tetrahedra with $d_1=3\lambda, d_2=3\lambda, d=7\lambda$. On the right hand side, we've plotted the case $d_1=9\lambda, d_2=3\lambda, d=21\lambda$.} \label{isoplot}
\end{figure}

This reproduces the result previously obtained in \cite{valentin}. We can easily check that this reduces to the previous equilateral case when $d_1=d_2=d$ and we further confirm it by numerical simulations. The plots in figure \ref{isoplot}  represents numerical simulations of an isosceles \sj-symbol minus the analytical formula (\ref{isoequa}). Like for the previous plot, we have multiplied the data by $\lambda^{5/2}$ to see how the coefficient of the NNLO order is approached and we have divided by $\cos (S_R+\pi/4)$ (NNLO oscillations) to suppress the oscillations.
Finally, the geometrical interpretation of the term $F(d,d_1,d_2)$ remains to be understood. If we can't provide it with a geometrical meaning, there is little hope to interpret the NLO coefficient $\Im(A(x_+))$ in the generic case. Nevertheless, we give a more compact expression for the denominator of $F$:
\be
96\left( 4d^2-d_1^2 \right)\left(4d^2-d_2^2 \right)\left( 4{d}^{2}-d_1^{
2}-d_2^{2} \right)=
\,
96^3\f{V^2}{\cos^2\theta}.
\ee
We still need to express the numerator of $F$ in term of geometrical objects. For instance, we could express it in term of $d^2$, $(4d^2-d_1^2)(4d^2-d_2^2)$ and $(4{d}^{2}-d_1^{2}-d_2^{2})$, which would provide a formula in term of the volume and the dihedral angles. Nevertheless, we haven't been able to find such a useful rewriting of this NLO coefficient.

%%%%%%%%%%%%%%%
\section*{Conclusion}
%%%%%%%%%%%%%%%

We investigated the asymptotical behavior of the \sj-symbol. Starting from its expression as a (finite) sum over (half-)integers of algebraic combinations of factorials, we followed the footsteps of \cite{razvan} and showed that one can derive systematically the corrections to the leading order formula at any order. The method relies on three steps. First, we use the Stirling formula (with the appropriate corrections) to approximate the factorials. Second, we consider the sum as a Riemann sum and approximate it by an integral (over the real line). Finally, we perform a saddle point approximation to compute the behavior of the \sj-symbol for (homogeneously) large spins.

Using this framework, we showed that we recover an oscillating leading order (LO) with frequency given by the Regge action as is well-known and was already proved in \cite{razvan}. Then we computed analytically the next-to-leading (NLO) corrections. The formula that we obtain is explicit, although not compact, and we could not interpret it geometrically in a clear way. Nevertheless, we performed two simple checks. First, we checked that our complicated formula reduces to the known expression for the NLO for isosceles tetrahedra \cite{valentin}. Second, we checked it numerically in various cases and found a perfect fit. These numerical simulations also confirmed that the NLO is a $\f\pi2$-phase shift with respect to the LO (the NLO is given by a $\sin$ instead of a $\cos$) and that the NNLO is back in phase with the LO (again a $\cos$), which confirms our expectation of an alternating asymptotical series in $\cos+\f1{j}\sin+\f1{j^2}\cos+\f1{j^3}\sin+\dots$.

We point out that we computed in details the corrections due to the Stirling formula and to the saddle point approximation. However we didn't study the Riemann sum approximation. It does not contribute to the LO and NLO. It will only enter at the level of the NNLO.

This work is mainly technical and can be applied to the computation of gravitational correlations for 3d quantum gravity following \cite{graviton, josh, valentin, 4d}. It will enter the quantum corrections to the propagator/correlations at second order, as was shown in \cite{josh}. Indeed, the first order corrections are derived from the path integral of the Regge action, while the deviations from the Regge action as computed here enter at second order (as two-loop corrections). From this perspective, this NLO of the \sj-symbol describes the leading order deviation of quantum gravity with respect to the classical gravity.

Beyond the technicality of the paper, our purpose was to show that computing such corrections is indeed possible (although it does lead to complicated expressions) and that similar methods could be used for 4d spinfoam gravity. Although these methods allow straightforward (but lengthy) analytical calculations, which might be handled by a computer program, their drawback is the loss of the  geometrical meaning of the expressions obtained. An alternative way to proceed is to use the exact recursion relations satisfied by the \sj-symbol (see \cite{SG}) and other spinfoam amplitudes (see \cite{morefun}) to probe the asymptotic behavior and the induced corrections of the correlations. This is left to future investigation \cite{next1}.

%%%%%%%%%%%%%%%
\section*{Ackowledgements}
%%%%%%%%%%%%%%%

The numerical simulations and plots were done using Mathematica 5.0.
MD and ER are partially supported by the ANR ``Programme Blanc" grant LQG-06.

%%%%%%%%%%%%%%%%%%%%%%%%%%%%%%%%%%%%%%%%%%%%%%%%%%
\appendix

%%%%%%
\section{The \sj-symbol - recoupling theory}
\label{6j}
%%%%%%

The \sj-symbol is a real number and it is obtained by combining four normalized Clebsh-Gordan coefficients along the six edges of a tetrahedron, with edge lengths given by $j_{IJ}+\frac{1}{2}=\frac{d_{j_{IJ}}}{2}$ ($0\leq I<J \leq 3$).
\begin{figure}[ht]
\begin{center}
\includegraphics[width=3cm]{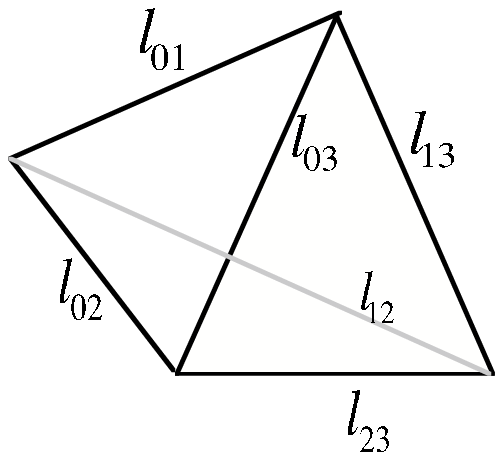}
\caption{Tetrahedron: edge lengths are given by $l_{IJ}=\frac{d_{j_{IJ}}}{2}$}
\end{center}
\end{figure}
We usually express the 6j-symbol in term of  the Wigner 3j-symbols~:
%which are equal to the Clebsh-Gordan coefficient up to a normalization factor~:
\equa{ \tabl{ll}{
 \left\{ \tabl{lll}{j_{01}&j_{02}&j_{03}\\
j_{23}&j_{13}&j_{12}\\} \right\}=& \sum_\alpha (-1)^{j_{01}+j_{03}+j_{01}-\alpha_{01}-\alpha_{03}-\alpha_{01}} \left( \tabl{lll}{j_{01}&j_{12}&j_{13}\\
\alpha_{01}&\alpha_{12}&-\alpha_{13}\\} \right)\\
& \left( \tabl{lll}{j_{13}&j_{23}&j_{03}\\
\alpha_{13}&\alpha_{23}&\alpha_{03}\\} \right) \left( \tabl{lll}{j_{03}&j_{02}&j_{01}\\
\alpha_{03}&\alpha_{02}&-\alpha_{01}\\} \right) \left( \tabl{lll}{j_{02}&j_{23}&j_{12}\\
\alpha_{02}&\alpha_{23}&\alpha_{12}\\} \right).
}}
The Wigner 3j symbols are very simply related to the Clebsh-Gordan coefficients $<j_{01} j_{12} \alpha_{01} \alpha_{12} | j_{13} \alpha_{13}>$ by:
$$<j_{01} j_{12} \alpha_{01} \alpha_{12} | j_{13} \alpha_{13}>= (-1)^{j_{01}-j_{12}+\alpha_{13}}(2j_{13}+1/2)^{1/2} \left( \tabl{lll}{j_{01}&j_{12}&j_{13}\\
\alpha_{01}&\alpha_{12}&-\alpha_{13}\\} \right),$$
And Racah gave a general formulae for the Clebsh-Gordan coefficient:
$$\tabl{ll}{ <j_{01} j_{12} \alpha_{01} \alpha_{12} | j_{13} \alpha_{13}>&= \delta( \alpha_{01}+ \alpha_{12}, \alpha_{13}) \Delta(j_{01}j_{12}j_{13}) \\
&\\
& \sqrt{(2j_{13}+1)(j_{01}+\alpha_{01})!(j_{01}-\alpha_{01})!(j_{12}+\alpha_{12})!(j_{12}-\alpha_{12})!(j_{13}+\alpha_{13})!(j_{13}-\alpha_{13})!}\\
&\\
&\sum_\mu \frac{(-1)^\mu}{(j_{01}-\alpha_{01}-\mu)!(j_{13}-j_{12}+\alpha_{01}+\alpha)!(j_{12}+\alpha_{12}-\mu)! (j_{13}-j_{01}-\alpha_{12}+\alpha)! \mu!(j_{01}+j_{12}-j_{13}-\mu)!}
}$$
where $\Delta(j_{01}, j_{12}, j_{13})= \frac{(j_{01}+j_{12}-j_{13})!(j_{01}-j_{12}+j_{13})!(-j_{01}+j_{12}+j_{13})!}{(j_{01}+j_{12}+j_{13}+1)!}$
From these, Racah gave a tensorial formulae for the 6j-symbol, the Racah's single sum formulae:
\equa{
\tabl{ll}{
\left\{ \tabl{lll}{j_{01}&j_{02}&j_{03}\\
j_{23}&j_{13}&j_{12}\\} \right\}= &\sqrt{\Delta(j_{01}, j_{02}, j_{03}) \Delta(j_{23}, j_{02}, j_{12}) \Delta(j_{23}, j_{13}, j_{03}) \Delta(j_{01}, j_{13}, j_{12}) }\\
& \displaystyle{\sum_{\textrm{max } v_I}^{\textrm{min } p_j}} (-1)^t \frac{(t+1)!}{\prod_i(t-v_I)! \prod_j (p_j-t)!}
}}
with $ v_K= \displaystyle{\sum_{I \neq K}} j_{IK} \; \forall K \in \{ 0, \cdots, 3\}$ and  $p_k= \displaystyle{\sum_{i \neq 0, k}} (j_{0i} + j_{ki}) \; \forall k \in \{1 \cdots 3 \}$.
%
% and $\Delta(j_{01}, j_{02}, j_{03})= %\frac{(j_{01}+j_{02}-j_{03})!(j_{01}-j_{02}+j_{03})!(-j_{01}+j_{02}+j_{03})!}{(j_{01}+j_{02}+j_{03}+1)!}$

%%%%%%%%%%%%%%%%%%%
\section{Factorials} \label{factorials}
%%%%%%%%%%%%%%%%%%%%

The factorial $n!$ is defined for a positive integer $n$ as:
$$
n! \equiv n(n-1) \cdots 2\cdot 1=\Gamma (n+1),
$$
where $\Gamma(n)$ is the gamma function for integers $n$. This definition is generalized to non-integer values. Using the identities for the $\Gamma$ function, we write explicitly the values for half-integers:
$$
(-\frac{1}{2})!=\sqrt{\pi}, \qquad
(\frac{1}{2})!= \frac{\sqrt{\pi}}{2},\qquad
(n-\frac{1}{2})!=\frac{\sqrt{\pi}}{2^n}(2n-1)!! =\frac{\sqrt{\pi}(2n)!}{2^{2n}n!},\qquad
(n+\frac{1}{2})!=\frac{\sqrt{\pi}}{2^{n+1}}(2n+1)!!=\frac{\sqrt{\pi}(2n+1)!}{2^{2n+1}n!},$$
%$$\tabl{l}{
%(-\frac{1}{2})!=\sqrt{\pi} \\
%\\
%(\frac{1}{2})!= \frac{1}{2}\sqrt{\pi}\\
%\\
%(n-\frac{1}{2})!=\frac{\sqrt{\pi}}{2^n}(2n-1)!! =\frac{\sqrt{\pi}(2n)!}{2^{2n}n!}\\
%\\
%(n+\frac{1}{2})!=\frac{\sqrt{\pi}}{2^{n+1}}(2n+1)!!=\frac{\sqrt{\pi}(2n+1)!}{2^{2n+1}n!} \\
%}$$
where $n!!$ is the double factorial~:
$$
n!! \equiv \left\{ \tabl{l}{
n \cdot (n-2) \cdots 5\cdot 3 \cdot 1 \;\; \textrm{ if } n>0 \textrm{ odd,} \\
n \cdot (n-2) \cdots 6\cdot 4 \cdot 2 \;\; \textrm{ if } n>0 \textrm{ even,} \\
1 \;\; \textrm{ if } n= -1 \textrm{ or } 0.
} \right .
$$
%Since
%$$ \tabl{ll}{
%(2n+1)!! 2^n n! &= [(2n+1)(2n-1) \cdots 1] [2n \cdot 2 (n-1) \cdot 2(n-2) \cdots  2 (1)]\\
%&=[(2n+1)(2n-1) \cdots 1] [2n \cdot (2n-2) \cdot (2n-4) \cdots  2]\\
%&= (2n+1)!
%}
%$$
%It follows that
%\equa{
%(2n+1)!! = \frac{(2n+1)!}{2^nn!}
%}
%Also, since
%$$\tabl{ll}{
%(2n-1)!! 2^n n! &= [(2n-1)(2n-3) \cdots 1] [2n \cdot 2 (n-1) \cdot 2(n-2) \cdots  2 (1)]\\
%&=[(2n-1)(2n-3) \cdots 1] [2n \cdot (2n-2) \cdot (2n-4) \cdots  2]\\
%&= (2n)! }$$
%It follows that
%\equa{
%(2n-1)!!=\frac{(2n)!}{2^nn!}
%}
%Therefore, we deduce that
%\equa{\tabl{l}{
%(n+\frac{1}{2})!=\frac{\sqrt{\pi}(2n+1)!}{2^{2n+1}n!}\\
%\\
%(n-\frac{1}{2})!=\frac{\sqrt{\pi}(2n)!}{2^{2n}n!}
%}}
Using the asymptotic expansion of a large factorial $n! \sim \sqrt{2 \pi n}\left(\frac{n}{e}\right)^n\left(1 + \frac{1}{12n}=\frac{1}{288n^3}-\frac{139}{51840n^3}-\frac{571}{2488320n^4}\cdots \right)$, we can get an asymptotic expansion for:
\equa{\tabl{l}{
(n+1/2)! \sim \sqrt{2 \pi} e^{(n+1)\ln(n)-n} \left(1+\frac{1}{2n}\right) \left(1+ \frac{11}{12(2n)} + \frac{1}{288(2n)^2}- \frac{139}{51840(2n)^3}-\frac{571}{2488320(2n)^4} + \cdots \right) \\
\;\;\;\;\;\;\;\;\;\;\;\;\;\;\;\;\;\;\;\;\;\;\;\;\left(1 - \frac{1}{12n}- \frac{1}{288n^2}+ \frac{139}{51840n^3}+\frac{571}{2488320n^4} - \cdots \right),\\
\\
(n-\frac{1}{2})! \sim \sqrt{2 \pi} e^{n\ln(n)-n}\left(1+ \frac{11}{12(2n)} + \frac{1}{288(2n)^2}- \frac{139}{51840(2n)^3}-\frac{571}{2488320(2n)^4} + \cdots \right) \\
\;\;\;\;\;\;\;\;\;\;\;\;\;\;\;\;\;\;\;\;\;\;\;\;\left(1 - \frac{1}{12n}- \frac{1}{288n^2}+ \frac{139}{51840n^3}+\frac{571}{2488320n^4} - \cdots \right),\\
}}
or more simply, at the next-to-leading order:
\equa{\tabl{l}{
n! \sim \sqrt{2 \pi n}\left(\frac{n}{e}\right)^n\left(1 + \frac{1}{12n}\right),\\
\\
(n+\frac{1}{2})! \sim \sqrt{2 \pi} e^{(n+1)\ln(n)-n}\left(1+ \frac{11}{24n}  \right), \\
\\
(n-\frac{1}{2})! \sim \sqrt{2 \pi} e^{n\ln(n)-n}\left(1- \frac{1}{24n}  \right). \\
}}

%%%%%%
\section{First approximation: factorials $\longrightarrow$ next to leading order of the Stirling formula}
\label{stirling}
%%%%%%

In this section, all computations are done at the next-to-leading order.
We replace the factorials in equation (\ref{Racah_dj}) by their respective asymptotic expansion. %(details concerning the asymptotic expansions of $(n+1/2)!$ and (n-1/2)! are given in appendix
%\ref{factorials}):
%\equa{ \label{approxstirling}
%\tabl{l}{
%n! \sim \sqrt{2 \pi }e^{(n+\frac{1}{2})\ln(n)-n}\left(1 + \frac{1}{12n}\right)\\
%\\
%(n+\frac{1}{2})! \sim \sqrt{2 \pi} e^{(n+1)\ln(n)-n}\left(1+ \frac{11}{24n}  \right) \\
%\\
%(n-\frac{1}{2})! \sim \sqrt{2 \pi} e^{n\ln(n)-n}\left(1- \frac{1}{24n}  \right). \\
%}}

\begin{itemize}
\item Then, a typical triangle coefficient:
$$\Delta(\lambda d_{j_{01}}, \lambda d_{j_{02}}, \lambda d_{j_{03}})= \frac{\left(\frac{\lambda }{2}(d_{j_{01}}+d_{j_{02}}-d_{j_{03}})-\frac{1}{2}\right)!\left(\frac{\lambda }{2}(d_{j_{01}}-d_{j_{02}}+d_{j_{03}})-\frac{1}{2}\right)!\left(\frac{\lambda }{2}(-d_{j_{01}}+d_{j_{02}}+d_{j_{03}})-\frac{1}{2}\right)!}{\left(\frac{\lambda }{2}(d_{j_{01}}+d_{j_{02}}+d_{j_{03}})-\frac{1}{2}\right)!}$$
will be
$$
\tabl{ll}{
\Delta(\lambda d_{j_{01}}, \lambda d_{j_{02}}, \lambda d_{j_{03}})&=2\pi [ e^{-\frac{\lambda }{2}(d_{j_{01}}+d_{j_{02}}+d_{j_{03}})\ln[\frac{\lambda }{2}(d_{j_{01}}+d_{j_{02}}+d_{j_{03}})]+\frac{\lambda }{2}(d_{j_{01}}+d_{j_{02}}+d_{j_{03}})}\left(1+\frac{1}{12\lambda (d_{j_{01}}+d_{j_{02}}+d_{j_{03}})}\right)\\
&\\
& \;\;\; e^{\frac{\lambda }{2}(d_{j_{01}}+d_{j_{02}}-d_{j_{03}})\ln[\frac{\lambda }{2}(d_{j_{01}}+d_{j_{02}}-d_{j_{03}})]-\frac{\lambda }{2}(d_{j_{01}}+d_{j_{02}}-d_{j_{03}})}\left(1-\frac{1}{12\lambda (d_{j_{01}}+d_{j_{02}}-d_{j_{03}})}\right)\\
&\\
& \;\;\; e^{\frac{\lambda }{2}(d_{j_{01}}-d_{j_{02}}+d_{j_{03}})\ln[\frac{\lambda }{2}(d_{j_{01}}-d_{j_{02}}+d_{j_{03}})]-\frac{\lambda }{2}(d_{j_{01}}-d_{j_{02}}+d_{j_{03}})}\left(1-\frac{1}{12\lambda (d_{j_{01}}-d_{j_{02}}+d_{j_{03}})}\right)\\
&\\
& \;\;\; e^{\frac{\lambda }{2}(-d_{j_{01}}+d_{j_{02}}+d_{j_{03}})\ln[\frac{\lambda }{2}(-d_{j_{01}}+d_{j_{02}}+d_{j_{03}})]-\frac{\lambda }{2}(-d_{j_{01}}+d_{j_{02}}+d_{j_{03}})}\left(1-\frac{1}{12\lambda (-d_{j_{01}}+d_{j_{02}}+d_{j_{03}})}\right)\\
}
$$
which simplifies
\equa{\tabl{ll}{
\Delta(\lambda d_{j_{01}}, \lambda d_{j_{02}}, \lambda d_{j_{03}})&=2\pi e^{\frac{\lambda }{2}[(-d_{j_{01}}+d_{j_{02}}+d_{j_{03}})\ln(-d_{j_{01}}+d_{j_{02}}+d_{j_{03}})+(d_{j_{01}}-d_{j_{02}}+d_{j_{03}})\ln(d_{j_{01}}-d_{j_{02}}+d_{j_{03}})]}\\
\\
&e^{-\frac{\lambda }{2}[(d_{j_{01}}+d_{j_{02}}-d_{j_{03}})\ln(d_{j_{01}}+d_{j_{02}}-d_{j_{03}})+(d_{j_{01}}+d_{j_{02}}+d_{j_{03}})\ln(d_{j_{01}}+d_{j_{02}}+d_{j_{03}})]}\\
\\
&[1-\frac{1}{12\lambda }(\frac{1}{-d_{j_{01}}+d_{j_{02}}+d_{j_{03}}}+\frac{1}{d_{j_{01}}-d-{j_{02}}+d_{j_{03}}}+\frac{1}{d_{j_{01}}+d_{j_{02}}-d-{j_{03}}} -\frac{1}{d_{j_{01}}+d_{j_{02}}+d_{j_{03}}})].
}}
The factor $\sqrt{\Delta(\lambda d_{j_{01}}, \lambda d_{j_{02}}, \lambda d_{j_{03}}) \Delta(\lambda d_{j_{23}}, \lambda d_{j_{02}}, \lambda d_{j_{12}}) \Delta(\lambda d_{j_{23}}, \lambda d_{j_{13}}, \lambda d_{j_{03}}) \Delta(\lambda d_{j_{01}},\lambda d_{j_{13}}, \lambda d_{j_{12}}) }$ in equation (\ref{Racah_dj}) can then easily be put into the form:
\equa{
(2\pi)^2e^{\frac{\lambda }{2}h(d_{j_{IJ}})}\left( 1-\frac{1}{24\lambda }H(d_{j_{IJ}}) \right)
}
where \equa{
\tabl{l}{
h(d_{j_{IJ}})= \displaystyle{\sum_{I<J}}d_{j_{IJ}}h_{d_{j_{IJ}}} \\
\textrm{with } h_{d_{j_{IJ}}}= \frac{1}{2} \ln \left( \frac{(d_{j_{IJ}}-d_{j_{IK}}+d_{j_{IL}})(d_{j_{IJ}}+d_{j_{IK}}-d_{j_{IL}})(d_{j_{IJ}}-d_{j_{JK}}+d_{j_{JL}})(d_{j_{IJ}}+d_{j_{JK}}-d-{j_{JL}})}{(d-{j_{IJ}}+d_{j_{IK}}+d_{j_{IL}})(-d_{j_{IJ}}+d_{j_{IK}}+d_{j_{IL}})(d_{j_{IJ}}+d_{j_{JK}}+d_{j_{JL}})(-d_{j_{IJ}}+d_{j_{JK}}+d_{j_{JL}})}\right)\\
 \;\;\; K \neq L \textrm{ and } K,L \neq I,J\\
H(d_{j_{IJ}})=2\displaystyle{\sum_{j,K}} \frac{1}{\tilde{p}_j-\tilde{v}_K} -2 \displaystyle{\sum_K} \frac{1}{\tilde{v}_K} \textrm{ where } K \in \{0, \cdots, 3 \} \textrm{ and } j \in \{1,\cdots, 3\} \\
}}
and we recall that  $ \tilde{v}_K= \displaystyle{\sum_{I \neq K}}\frac{ d_{j_{IK}}}{2} \; \forall K \in \{ 0, \cdots, 3\}, \; \tilde{p}_k= \displaystyle{\sum_{i \neq 0, k}} \frac{(d_{j_{0i}} + d_{j_{ki}})}{2} \; \forall k \in \{1 \cdots 3 \}$.
\item We now replace the factorials in the sum of (\ref{Racah_dj}) by their approximations and we change of variables: $t=\lambda x$:
\equa{\tabl{ll}{
\Sigma(\lambda d_{j_{IJ}})&= \displaystyle{\sum_{x=\textrm{max }\tilde{ v}_I}^{\textrm{min } \tilde{p}_j}} (-1)^{\lambda x} \frac{(\lambda x+1)(\lambda x)! \prod_j(\lambda (\tilde{p_j}-x)) \prod_j (\lambda (\tilde{p}_j-x)-1)}{\prod_I (\lambda (x-\tilde{v}_I)+3/2)(\lambda (x-\tilde{v}_I)+1/2)!\prod_j(\lambda (\tilde{p}_j-x))!}\\
&= \frac{1}{(2\pi)^3} \displaystyle{\sum_{x=\textrm{max } v_I}^{\textrm{min } p_j}} e^{G_1(x)}G_2(x)
}}
where \equa{\tabl{ll}{
G_1(x)=& i\pi \lambda x + 3 \ln \lambda  + \ln x +2 \displaystyle{\sum_j} \ln (\tilde{p}_j-x) -\displaystyle{\sum_I} \ln (x- \tilde{v}_I)+(\lambda x+1/2)(\ln x + \ln \lambda ) -\lambda x+ \displaystyle{\sum_I} \lambda (x-\tilde{v}_I) \\
&- \displaystyle{\sum_I} (\lambda (x-\tilde{v}_I)+1)(\ln \lambda  + \ln (x-\tilde{v}_I))-\displaystyle{\sum_j} (\lambda (\tilde{p}_j-x)+1/2)(\ln \lambda  +\ln(\tilde{p}_j-x)) +\displaystyle{\sum_j} \lambda (\tilde{p}_j-x)
 }}
which can be simplified using the fact that $\sum_I \tilde{v}_I= \sum_j \tilde{p}_j$:
\equa{
G_1(x)= -2\ln \lambda +\frac{1}{2}\ln\frac{x^3\prod_j(\tilde{p}_j-x)^3}{\prod_I(x-\tilde{v}_I)^4} +\lambda \left[ i\pi x  +x \ln x - \sum_I (x-v_I)  \ln (x -v_I) - \sum_j (p_j-x)\ln(p_j-x) \right]
}
and
\equa{\tabl{l}{
G_2(x)= \frac{1+\frac{1}{12\lambda x}}{(1+\frac{3}{2\lambda (x-\tilde{v}_I)}\prod_I(1+\frac{11}{24\lambda (x-\tilde{v}_I)})\prod_j(1+\frac{1}{12\lambda (p_j-x)})}\\
\;\;\;\;\;\;= 1-\frac{1}{\lambda }\left(- \frac{13}{12x}+ \displaystyle{\sum_I}\frac{47}{24(x-v_I)} + \sum_j \frac{13}{12(p_j-x)} +O\left( \frac{1}{\lambda }\right)\right)
}}
Moreover,
\equa{
e^{G_1(x)}=\frac{1}{\lambda ^2}e^{F(x)+\lambda f(x)}}
where \equa{\tabl{l}{
f(x)= i \pi x+ x \ln(x) - \displaystyle{\sum_I}(x-v_K) \ln(x-v_I) - \displaystyle{\sum_j}(p_j-x) \ln(p_j-x) \\
F(x)= \frac{1}{2} \ln \left( \frac{x^3\prod_j(p_j-x)^3}{\prod_I (x-v_I)^4 } \right).
}}
Then the sum can be approximated by:
\equa{
\Sigma(\lambda d_{j_{IJ}})=\frac{1}{(2\pi)^3\lambda ^2}\displaystyle{\sum_{x=\textrm{max } v_I}^{\textrm{min } p_j}} e^{\lambda f(x)+F(x)} \left( 1 -\frac{1}{12\lambda }G(x) + O\left(\frac{1}{\lambda }\right)\right) e^{\lambda f(x)}
}
where \equa{
G(x)=-\frac{13}{x}+ \displaystyle{\sum_K}\frac{47}{24(x-v_K)} + \sum_j \frac{13}{p_j-x}. \\
}
\end{itemize}

%%%%%%%%%%%%%%%%%%%
%\section{Second approximation: Riemann sum $\longrightarrow$ integral } \label{riemann}

%%%%%%%%%%%%%%%%%%%%
\section{Third approximation: the stationary phase method} \label{saddlepoint}
We are interested in the $1/\lambda  $ expansion of the integral:
$$I=\displaystyle{\int_{\textrm{max } \tilde{v}_I/2}^{\textrm{min } \tilde{p}_j/2}} dx e^{F(x)} \left( 1 -\frac{1}{12\lambda }G(x) + O\left(\frac{1}{\lambda }\right)\right) e^{\lambda f(x)}.$$
We do not give here the proof of the whole expansion (equation (\ref{complete})) because of the heavy formalism but we directly prove the next to leading order formula (equation (\ref{intNLO})); the procedure is the same but the computations are easier.
The asymptotic expansion of such an integral is given by contributions around the stationary points of the phase denoted $x_0$. We expand the phase $f(x)$ around the stationary points $x_0$ at fourth order and the function $g(x)=e^{F(x)}\left( 1- \frac{1}{12\lambda }G(x) \right)$ at second order and we extend the integration to infinity.
\equa{
\tabl{ll}{
I \sim \displaystyle{\sum_{x_0}} \displaystyle{ \int_{-\infty}^{+ \infty} }& d(\delta x) \left(g(x_0) + g^\prime(x_0) \delta x + \frac{1}{2} g^{\prime \prime}(x_0) (\delta x)^2 \right) e^{\lambda \left( f(x_0)+ \frac{1}{2}f^{\prime \prime}(x_0) (\delta x)^2\right)} \\
& \left(1+ \lambda  \left(\frac{1}{3!} f^{(3)}(x_0) (\delta x)^3 + \frac{1}{4!}f^{(4)}(x_0) (\delta x)^4\right) +\frac{\lambda ^2}{2} \left( \frac{1}{3!} f^{(3)}(x_0) (\delta x)^3 \right)^2  +O(\lambda ^2)\right)
}}
where in our case, $g(x)= e^{F(x)}\left( 1- \frac{1}{12\lambda }G(x) \right)$ and then the integration are "generalized" Gaussians:
\equa{
\tabl{ll}{
I \sim \displaystyle{\sum_{x_0}}  e^{F(x_0)+\lambda f(x_0)} &[ \left( 1 -\frac{1}{12\lambda }G(x_0) \right) \displaystyle{ \int_{-\infty}^{+ \infty} } d(\delta x) e^{-\lambda (\frac{-f^{\prime \prime}(x_0)}{2})(\delta x)^2} \\
&+\frac{1}{2}\left( (F^\prime(x_0))^2+ F^{\prime \prime} (x_0) \right) \displaystyle{ \int_{-\infty}^{+ \infty} }d(\delta x) (\delta x)^2 e^{-\lambda (\frac{-f^{\prime \prime}(x_0)}{2})(\delta x)^2} \\
&+ \lambda  \left(\frac{f^{(4)}(x_0)}{4!} + \frac{f^{(3)}(x_0)}{3!}F^\prime(x_0)\right) \displaystyle{ \int_{-\infty}^{+ \infty} }d(\delta x) (\delta x)^4 e^{-\lambda (\frac{-f^{\prime \prime}(x_0)}{2})(\delta x)^2} \\
&+ \frac{\lambda ^2}{2} \left( \frac{f^{(3)}(x_0)}{3!} \right)^2 \displaystyle{ \int_{-\infty}^{+ \infty} }d(\delta x) (\delta x)^6 e^{-\lambda (\frac{-f^{\prime \prime}(x_0)}{2})(\delta x)^2}+ O \left(\frac{1}{\lambda ^{3/2}} \right) ]
}}
which can easily be computed:
\equa{\tabl{ll}{
I \sim \displaystyle{\sum_{x_0}} &\sqrt{\frac{2\pi}{-f^{\prime \prime}(x_0)\lambda } } e^{F(x_0)+\lambda f(x_0)} \\
&\left[1 +
\frac{1}{\lambda } \left( -\frac{G(x_0)}{12}- \frac{F^{\prime \prime}(x_0)+(F^{\prime}(x_0))^2}{2 f^{\prime \prime}(x_0)}+\frac{f^{(4)}(x_0)+4f^{(3)}(x_0)F^\prime(x_0)}{8(f^{\prime \prime}(x_0))^2}- \frac{5(f^{(3)}(x_0))^2}{24(f^{\prime \prime}(x_0))^3} \right) + O \left(\frac{1}{\lambda } \right) \right]
}}

%%%%%%%%%%%%%%%%%%%%%%%%%%%%%%%%%%%%%%%%%%%%%%%%%%%

\end{document}